\documentclass[preprint2]{aastex}

\shorttitle{Low-Ionization Emission Regions in Quasars}
\shortauthors{Matsuoka et al.}

\begin{document}


\title{Low-Ionization Emission Regions in Quasars: \\
  Gas Properties Probed with Broad \ion{O}{1} and \ion{Ca}{2} Lines}

\author{Y. Matsuoka\altaffilmark{1,}\altaffilmark{3}}
\author{K. Kawara\altaffilmark{1}}
\and
\author{S. Oyabu\altaffilmark{2}}


\altaffiltext{1}{Institute of Astronomy, The University of Tokyo, 2-21-1, Osawa,  Mitaka, Tokyo 181-0015, 
  Japan; matsuoka@ioa.s.u-tokyo.ac.jp.}
\altaffiltext{2}{Institute of Space and Astronautical Science, Japan Aerospace Exploration Agency, 3-1-1,
  Yoshinodai, Sagamihara, Kanagawa 229-8510, Japan.}
\altaffiltext{3}{Research Fellow of the Japan Society for the Promotion of Science.}


\begin{abstract}
We have compiled the emission-line fluxes of \ion{O}{1} $\lambda$8446, \ion{O}{1} $\lambda$11287,
and the near-infrared (IR) \ion{Ca}{2} triplet ($\lambda$8579) observed in 11 quasars.
These lines are considered to emerge from the same gas as do the \ion{Fe}{2} lines in the low-ionized
portion of the broad emission line region (BELR).
The compiled quasars are distributed over wide ranges of redshift (0.06 $\le z \le$ 1.08) and of 
luminosity ($-29.8 \le M_{B} \le -22.1$), thus providing a useful sample to investigate the
line-emitting gas properties in various quasar environments.
The measured line strengths and velocities, as functions of the quasar properties, are analyzed using 
photoionization model calculations. 
We found that the flux ratio between the \ion{Ca}{2} triplet and \ion{O}{1} $\lambda$8446 is hardly dependent 
on the redshift or luminosity, indicating similar gas densities in the emission region from quasar to quasar.
On the other hand, a scatter of the \ion{O}{1} $\lambda$11287/$\lambda$8446 ratios appears to imply the diversity 
of the ionization parameter.
These facts invoke a picture of the line-emitting gases in quasars that have similar densities and are 
located at regions exposed to various ionizing radiation fluxes.
The observed \ion{O}{1} line widths are found to be remarkably similar over more than 3 orders
of magnitude in luminosity, which indicates a kinematically determined location of the emission region and is
in clear contrast to the case of \ion{H}{1} lines.
We also argue about the dust presence in the emission region since the region is suggested to be located 
near the dust sublimation point at the outer edge of the BELR.
\end{abstract}


\keywords{galaxies: active ---  galaxies: evolution --- galaxies: nuclei --- line: formation --- quasars: emission lines}



\section{Introduction}

Active galactic nuclei (AGNs) are known to have strong emission lines of various ion species.
Among them, the
\ion{Fe}{2} emission lines are one of the most prominent features in the ultraviolet (UV) to optical spectrum of many AGNs.
They have long been hoped to provide significant information about some aspects of the AGNs and their host environments,
e.g., energy budget of the line emission region 
and the epoch of the first star formation in the host galaxies.
The determination of the first star formation epoch is based on the standard theory that the iron enrichment in galaxies 
is delayed compared to 
that of the $\alpha$-elements, such as magnesium, due to their different origins; Type Ia supernovae for iron and Type II
supernovae for the $\alpha$-elements \citep{hamann93,yoshii98}.
The delay corresponds to the difference in life-times of the progenitors of the two types of supernovae, and is estimated to
be 0.3 -- 1 Gyr depending on the host galaxy environments \citep{yoshii96, matteucci01}.
Many observations have been devoted to the measurement of \ion{Fe}{2}/\ion{Mg}{2} line flux ratios in high-redshift quasars 
for this purpose over the last decade \citep[e.g.,][]{elston94,kawara96,dietrich02,dietrich03,iwamuro02,iwamuro04,freudling03,maiolino03}.
However, the observed \ion{Fe}{2}/\ion{Mg}{2} ratios show a large scatter, preventing a detection of any significant trend in the 
Fe abundance as a function of redshift.
While a part of the scatter might be due to the difference in the intrinsic Fe/Mg abundance ratio, it is presumable that 
the diversity of the physical condition within the line-forming gas, affecting line emissivities, is the main cause
\citep{verner03, baldwin04}.
In the same sense, a change of the observed \ion{Fe}{2}/\ion{Mg}{2} ratio as a function of redshift, if found, should 
be carefully examined to tell whether it reflects the abundance evolution or the systematic variation of the line emissivity.
Thus establishment of a method to probe the line-emitting gas and estimate its physical parameters such as
density and incident-ionizing radiation flux has been much awaited.

Unfortunately, the Fe$^+$ atom is characterized by an enormous numbers of possible electronic transitions, yielding the ``\ion{Fe}{2} 
pseudocontinuum'' often observed in AGN spectra, which makes analysis of the observations extremely difficult
from both the observational and theoretical viewpoints \citep[e.g., ][hereafter T06]{tsuzuki06}.
On the other hand, a promising approach is to use the emission lines emitted by simple atoms in the same region as the \ion{Fe}{2}
lines.
The most potent lines are \ion{O}{1} and \ion{Ca}{2}, whose co-spatial emergence with \ion{Fe}{2} is indicated by a resemblance of
their profiles \citep{rodriguez02a} and by a correlation between the line strengths \citep{persson88}.
Note that it is a natural consequence of similar ionization potentials of the relevant ions, i.e., 16.2 eV for \ion{Fe}{2}, 13.6 eV
for \ion{O}{1}, and 11.9 eV for \ion{Ca}{2}.

The first extensive study of the physical properties of \ion{O}{1} emitting gases in AGNs was presented by \citet{grandi80}, who 
observed the strongest \ion{O}{1} line, $\lambda$8446, as well as other weaker \ion{O}{1} lines in Seyfert 1 (Sy1) galaxies.
He found that \ion{O}{1} $\lambda$8446 lacks the narrow component that characterizes other permitted lines, and concluded that the
line is purely a BELR phenomenon.
He also suggested that \ion{O}{1} $\lambda$8446 is produced by Ly$\beta$ fluorescence, which was later confirmed by the observation
of I Zw 1, the prototype narrow-line Seyfert 1 (NLS1), by \citet{rudy89}.
\citet{rodriguez02b} compiled the UV and near-IR \ion{O}{1} lines, namely, $\lambda$1304, $\lambda$8446, and $\lambda$11287, in 
normal Sy1s and NLS1s in order to investigate their flux ratios.
They found that there must be an additional excitation mechanism for \ion{O}{1} $\lambda$8446---besides Ly$\beta$ fluorescence---which
they concluded is collisional excitation.
As for the \ion{Ca}{2} lines, extensive studies of Sy1s were presented by \citet{persson88} and \citet{ferland89}.
We have observed seven quasars at redshifts up to $\sim$1.0 for the purpose of obtaining the UV and near-IR \ion{O}{1} and \ion{Ca}{2} lines, thus extending 
the previous studies to include quasars at high redshifts.
Photoionization model calculations were performed and compared with the observations, which led us to conclude that the \ion{O}{1} and
\ion{Ca}{2} lines are formed in a gas with density $n_{\rm H} = 10^{11.5} - 10^{12.0}$ cm$^{-3}$, illuminated by the ionizing
radiation corresponding to the ionization parameter of $U = 10^{-3.0} - 10^{-2.5}$ \citep[][the latter is referred to as Paper I hereafter]{matsuoka05, 
matsuoka07}.

Now that the general properties of the \ion{O}{1} and \ion{Ca}{2} emitting gas, as a whole, are thus being revealed, we should turn our
eyes to the following question: how are these gas properties related to the quasar characteristics, such as redshift and luminosity?
Here we present results of a compilation of the emission-line fluxes of \ion{O}{1} $\lambda$8446, \ion{O}{1} $\lambda$11287, and the near-IR 
\ion{Ca}{2} triplet ($\lambda$8498, $\lambda$8542, and $\lambda$8662) observed in 11 quasars.
The quasars are distributed over wide ranges of redshift (0.06 $\le z \le$ 1.08) and of luminosity ($-29.8 \le M_{B}
\le -22.1$), thus providing a useful sample to track the line-emitting gas properties in various quasar environments.
The observational and theoretical data used in this work are described in \S \ref{sec:data}, and results and discussion appear in \S
\ref{sec:results}.
Our conclusions are summarized in \S \ref{sec:summary}.

\section{Data \label{sec:data}}


\subsection{Observations}

A quasar sample is compiled from \citet{matsuoka05}, \citet{riffel06}, and Paper I. 
Their observational characteristics\footnote{
The $B$-band absolute magnitudes of the quasars are taken from \citet{veron06} who assumed the cosmological parameters of 
$H_{\rm 0}$ = 50 km s$^{-1}$ Mpc$^{-1}$ and $q_{\rm 0}$ = 0.
In a flat cosmology with $H_{\rm 0}$ = 71 km s$^{-1}$ Mpc$^{-1}$, ${\rm \Omega}_{\rm M}$ = 0.29, and ${\rm \Omega}_{\rm \Lambda}$ = 0.71
\citep[e.g.,][]{riess04}, 
the absolute magnitudes would be systematically smaller than listed in Table \ref{sample} by 0.6 -- 0.7 mag
\citep[see Fig. 2 in][]{veron06}.
However, such a systematic change does not affect our arguments below.}
are summarized in Table \ref{sample}.
The sample in \citet{riffel06} mainly provides quasars in the local universe ($z \la$ 0.1);
they selected the quasars from the Palomer bright quasar survey of \citet{schmidt83} with the criteria of (1) 
being well-known
studied sources in the optical/UV and X-ray regions, and (2) having the $K$-band magnitude limited to $K <$ 12.
On the other hand, the quasars in \citet{matsuoka05} and Paper I constitute a high-redshift sample with redshifts up to $\sim$1.0.
The selection criteria used in those works are that (1) the UV spectrum is available in the {\it Hubble Space Telescope} Faint Object Spectrograph archive
\citep{evans04}; and (2) \ion{O}{1} $\lambda$8446, \ion{O}{1} $\lambda$11287, and the near-IR \ion{Ca}{2} triplet fall within the
atmospheric windows.
Then the brightest objects in the target list were observed, which have the $H$-band magnitudes of $H <$ 15.

\begin{figure}
\epsscale{.80}
\plotone{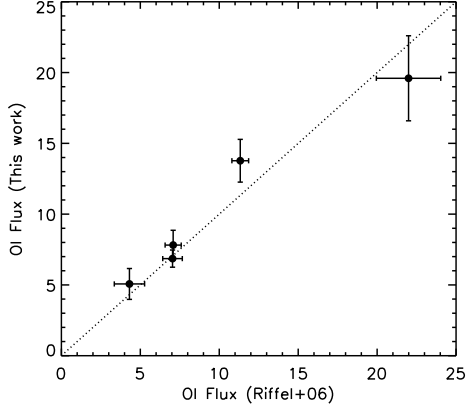}
\caption{The fluxes of \ion{O}{1} $\lambda$11287 measured in this work are plotted versus those listed in \citet{riffel06}.
  The units of both axes are 10$^{-15}$ ergs cm$^{-2} $ s$^{-1}$.
  The fluxes measured in PG 1126$-$041 (at the top-right corner) was multiplied by 0.5 in both axes in order to improve the 
  visibility.
  A dotted line represents the locations where our \ion{O}{1} fluxes are identical to those by \citet{riffel06}.
\label{compare}}
\end{figure}

Although the fluxes of the relevant \ion{O}{1} and \ion{Ca}{2} lines have been measured in \citet{riffel06}, we re-measured them
with the identical method to Paper I in order to remove any systematic differences due to the different
measurement strategies, especially for deblending the \ion{O}{1} $\lambda$8446 + the \ion{Ca}{2} triplet feature.
The 1$\sigma$ error of the spectra is assumed to be 1\% of the continuum level while their spectra apparently have the higher qualities.
The measured fluxes of \ion{O}{1} $\lambda$11287 is plotted versus those listed in \citet{riffel06} in Figure \ref{compare};
they are in good agreement with each other, which is well within the accuracy needed in this work.
We also re-measured the emission-line fluxes of PG 1116$+$215 studied in \citet{matsuoka05}, since we did not cover the \ion{Ca}{2}
lines in that work.

The measured line strengths are listed in Table \ref{lineflux} as the photon number flux ratios of the 
\ion{Ca}{2} triplet/\ion{O}{1} $\lambda$8446
and of \ion{O}{1} $\lambda$11287/\ion{O}{1} $\lambda$8446, and as the rest-frame equivalent widths (EWs) of \ion{O}{1} $\lambda$8446.
These forms of expression are particularly useful in discussing the line formation processes.
Hereafter they are expressed as $n$(\ion{Ca}{2})/$n$(\ion{O}{1} $\lambda$8446), \ion{O}{1} $n$($\lambda$11287)/$n$($\lambda$8446), 
and EW (\ion{O}{1} $\lambda$8446), respectively.

\subsection{Model Calculations\label{data_model}}

In order to interpret the observations, we utilize the model calculations presented in Paper I.
They are briefly summarized below.

The model calculations were performed in the framework of the photoionized BELR gas using the photoionization code Cloudy,
version 06.02 \citep{ferland98}.
Two shapes of the incident continuum were adopted, which are ``standard AGN ionizing continuum''s reported in T06
and in \citet[][hereafter K97]{korista97}.
However, we found little effect on the predicted \ion{O}{1} and \ion{Ca}{2} line ratios by changing 
the incident continuum shape from one to another.
The BELR gas was modeled to have a constant hydrogen density $n_{\rm H}$ (cm$^{-3}$) and be exposed to the ionizing continuum with a 
photon flux of $\Phi$ (s$^{-1}$ cm$^{-2}$).
In the calculations, the ionizing continuum flux was expressed with the ionization parameter $U \equiv \Phi/(n_{\rm H}c)$
where $c$ is the speed of light.
The gas column density was set to 10$^{23}$ cm$^{-2}$ and was changed in a range of $N_{\rm H} = 10^{17} - 10^{25}$ cm$^{-2}$, while
the chemical abundance was assumed to be solar.
Three cases of the microturbulence were considered, whose velocity $v_{\rm turb}$ is 0, 10, and 100 km s$^{-1}$.

The different continuum shapes and the microturbulent velocities were combined into four baseline models (Model 1 with T06 continuum and
$v_{\rm turb}$ = 0 km s$^{-1}$, Model 2 with T06 continuum and $v_{\rm turb}$ = 10 km s$^{-1}$, Model 3 with T06 continuum and 
$v_{\rm turb}$ = 100 km s$^{-1}$, and  Model 4 with K97 continuum and $v_{\rm turb}$ = 100 km s$^{-1}$).
We searched in the gas physical parameter space of ($n_{\rm H}$, $U$) in each model for the parameter sets that reproduce the
observed \ion{O}{1} and \ion{Ca}{2} line strengths, and
found that all baseline models with ($n_{\rm H}$, $U$) around ($10^{11.5}$ cm$^{-3}$, $10^{-3.0}$) best fit to the observations.
On the other hand, the observed shape and EW of the \ion{Fe}{2} UV emission could not be reproduced unless the microturbulent velocity of
$v_{\rm turb} >$ 100 km s$^{-1}$ was assumed \citep{baldwin04}.
Thus we adopt Model 3 as the default baseline model below, while we get little change of the results in this work concerning the 
analysis of the \ion{O}{1} and \ion{Ca}{2} lines when other baseline models are adopted.
The ($n_{\rm H}$, $U$) parameters that best fit to the observations in Model 3 are ($n_{\rm H}$, $U$) = (10$^{12.0}$ cm$^{-3}$, 10$^{-2.5}$),
which are used as the reference grid point below.




\section{Results and Discussion \label{sec:results}}

\subsection{A Picture of the Dust-Free Emission Region}

\begin{figure}
\epsscale{1.0}
\plotone{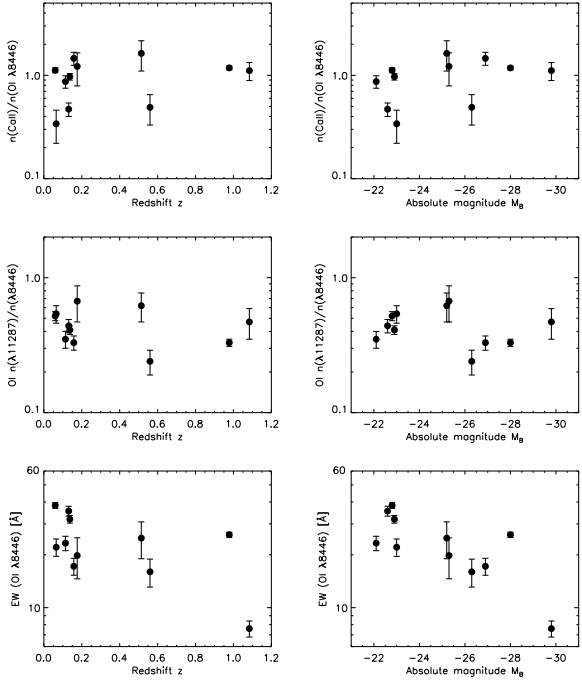}
\caption{The observed values of $n$(\ion{Ca}{2})/$n$(\ion{O}{1} $\lambda$8446), \ion{O}{1} $n$($\lambda$11287)/$n$($\lambda$8446),
  and EW (\ion{O}{1} $\lambda$8446) are plotted as functions of the redshift ({\it left}) and of the $B$-band absolute 
  magnitude ({\it right}). 
  \label{vsprop2}}
\end{figure}

\begin{figure}
\epsscale{1.00}
\plotone{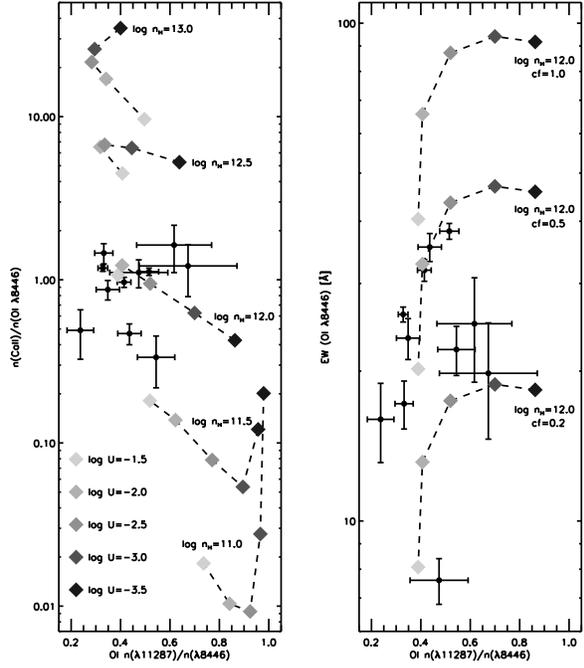}
\caption{Observed and theoretical values of $n$(\ion{Ca}{2})/$n$(\ion{O}{1} $\lambda$8446) ({\it left}) and EW (\ion{O}{1} $\lambda$8446) 
  ({\it right}) versus \ion{O}{1} $n$($\lambda$11287)/$n$($\lambda$8446).
  Filled circles represent the observations while diamonds represent the model predictions.
  The models with the same gas density, log $n_{\rm H}$ = 11.0, 11.5, 12.0, 12.5, and 13.0 ({\it left}), or with the same covering factor, 
  cf = 0.2, 0.5, and 1.0 with log $n_{\rm H}$ = 12.0 ({\it right}), are connected with dashed lines.
  Gray scale filling the diamonds distinguishes different values of the ionization parameter $U$ as indicated at the bottom-left corner
  of the left panel.
\label{oioi}}
\end{figure}

We plot the observed values of $n$(\ion{Ca}{2})/$n$(\ion{O}{1} $\lambda$8446), \ion{O}{1} $n$($\lambda$11287)/$n$($\lambda$8446),
and EW (\ion{O}{1} $\lambda$8446) as functions of the redshift and of the $B$-band absolute 
magnitude $M_{B}$ in Figure \ref{vsprop2}.
One of the remarkable results is found in the top panels; the \ion{Ca}{2}/\ion{O}{1} $\lambda$8446 ratio is hardly dependent on 
redshift or luminosity over the plotted range,
while the ratio is predicted to be very sensitive to the density of the line-emitting gas in the photoionization models.
We show the model predictions on the $n$(\ion{Ca}{2})/$n$(\ion{O}{1} $\lambda$8446) -- 
\ion{O}{1} $n$($\lambda$11287)/$n$($\lambda$8446) plane in Figure \ref{oioi} ({\it left}), as well as the observed values (see also Fig. 7 
in Paper I\footnote{
Note that Figure 7 in Paper I shows the predictions of Model 1, which is not adopted in this paper since the assumed microturbulent
velocity, $v_{\rm turb}$ = 0 km s$^{-1}$, could not reproduce the observed \ion{Fe}{2} UV emissions (see \S \ref{data_model}).
However, Model 3 adopted in this work predict very similar results to those shown in Figure 7 regarding the \ion{O}{1} and \ion{Ca}{2} 
emissions, while the whole pattern of contour is slightly ($\sim$ 0.5 dex) shifted to the high-density regime; the best-fit
parameters in Model 1 are ($n_{\rm H}$, $U$) = (10$^{11.5}$ cm$^{-3}$, 10$^{-3.0}$).
}).
The gas density $n_{\rm H}$ and the ionization parameter $U$ are changed around the reference grid point, ($n_{\rm H}$, $U$)
= (10$^{12.0}$ cm$^{-3}$, 10$^{-2.5}$), over 2 orders of magnitude in both parameters.
It is clearly seen that the predicted $n$(\ion{Ca}{2})/$n$(\ion{O}{1} $\lambda$8446) ratio increases monotonically with 
the increased gas density, and that all the observed values are marked with the density in the vicinity of the 
reference point, log $n_{\rm H}$ = 12.0.
Thus the similar density of the line-emitting gases are strongly indicated for the quasars distributed over these redshift and luminosity ranges.

The values of EW (\ion{O}{1} $\lambda$8446), both observed and calculated with the density of log $n_{\rm H}$ = 12.0, are shown in Figure \ref{oioi} 
({\it right}) as a function of the \ion{O}{1} $n$($\lambda$11287)/ $n$($\lambda$8446) ratio.
It shows that the predictions of EWs are also consistent with the observed data when log $n_{\rm H}$ = 12.0 and a covering fraction (cf) of the
line-emitting gas as seen from the central energy source of 0.2 -- 0.5 are assumed.
Note that the covering fraction could be much smaller if we assumed oxygen overabundance relative to the solar value.

On the other hand, a scatter of the observed data in the \ion{O}{1} $n$($\lambda$11287)/ $n$($\lambda$8446) axis seems to be related to the 
diversity of the ionization parameter (Fig. \ref{oioi}, {\it left}).
Note that the diversity of other parameters, such as microturbulent velocity, gas column density, and chemical composition,
could not explain this diagram since they significantly alter the $n$(\ion{Ca}{2})/$n$(\ion{O}{1} $\lambda$8446) ratio,
rather than \ion{O}{1} $n$($\lambda$11287)/ $n$($\lambda$8446), and thus they are unable to explain the observed similarity of the former 
ratios (Paper I).
It is also quite unlikely that the diversity of these parameter values is balanced out by the fine-tuned density in such a way that
the $n$(\ion{Ca}{2})/$n$(\ion{O}{1} $\lambda$8446) ratio is always kept to be $\sim$1.0, unless these lines are the dominant 
heating or cooling sources of the emission region.
As with the $n$(\ion{Ca}{2})/$n$(\ion{O}{1} $\lambda$8446) ratio, \ion{O}{1} $n$($\lambda$11287)/ $n$($\lambda$8446) is not clearly
dependent on the redshift or luminosity (Fig. \ref{vsprop2}, {\it middle left and middle right}).

The above arguments invoke a picture of the line-emitting gases in quasars that have similar densities and are 
located at regions exposed to various ionizing radiation fluxes.
It would be a consequence of the difference in distance to the central continuum source and/or in the intrinsic 
luminosity of the quasars.
Note that it is in clear contrast to the well-studied case of H$\beta$, whose emission regions in AGNs are known to be
characterized by similar ionization parameters.
In fact, reverberation mapping results for H$\beta$ show the emission region size ($r$) -- luminosity ($L$) relation of 
$r \propto L^{0.5}$, which is consistent with the constant ionization parameter regime \citep{peterson02, bentz06}.
Such a situation has long been expected in order to account for the remarkably similar AGN spectra over a broad range of luminosity,
and was incorporated into the locally optimally emitting cloud (LOC) model suggested by \citet{baldwin95}, that is, that the BELR is composed
of gas with widely distributed physical parameters and each emission line arises from its preferable environment.
On the other hand, the case in \ion{O}{1} and \ion{Ca}{2} lines apparently indicates that the location of the emission region is not 
radiation-selected.

\begin{figure}
\epsscale{0.80}
\plotone{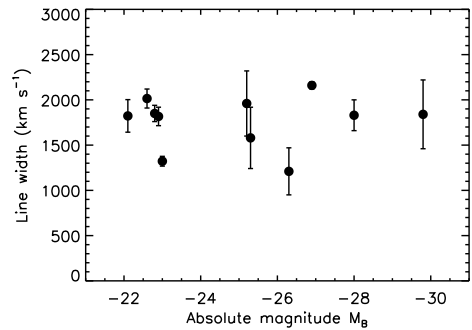}
\caption{The observed line widths of \ion{O}{1} $\lambda$11287 versus the $B$-band absolute magnitudes.
\label{mag_width}}
\end{figure}

In line with the above arguments, we found a clear difference of the velocity-luminosity relation between \ion{O}{1} and H$\beta$;
the measured \ion{O}{1} line widths are plotted versus $M_B$ in Figure \ref{mag_width}, which shows that the \ion{O}{1} line widths
are remarkably similar, concentrated around 1500 -- 2000 km s$^{-1}$, over more than 3 orders of magnitude in the $B$-band luminosity.
On the other hand, those for the hydrogen Balmer lines usually have a large scatter as shown by, e.g., \citet{kaspi00}; their sample of 
34 AGNs, spanning over 4 orders 
of magnitude in continuum luminosity, has the line widths of 1000 -- 10,000 km s$^{-1}$.
Such a trend is also indicated by \citet{persson88}, who reported that while the correlation between FWHM (\ion{O}{1}) and FWHM (H$\beta$) is good for
the small FWHM regime, the \ion{O}{1} lines grow systematically narrower than H$\beta$ at large line width.
\citet{rodriguez02a} conducted a detailed study of the near-IR emission line profiles in NLS1s, and found that \ion{O}{1}, \ion{Ca}{2}, and \ion{Fe}{2} lines
are systematically narrower than the broad components of other low-ionization lines such as hydrogen Paschen lines and \ion{He}{1} $\lambda$10830.
They also argued that these lines are produced in the outermost portion of the BELR, since their widths are just slightly
broader than those of [\ion{S}{3}] $\lambda$9531 which they assumed is formed in the inner portion of the narrow emission line region (NELR).
While the scattered widths of \ion{H}{1} lines could be interpreted as a consequence of the radiation-selected locations of the emitting gases, regardless of the
gas kinematics, the remarkable similarity of the \ion{O}{1} line widths might imply the kinematically determined emission regions.
In such a situation, the diversity of the ionization parameters, as discussed above, would be an inevitable result.
As stated by \citet{persson88}, there is clearly interesting information which could be deduced from studies of the \ion{O}{1} and \ion{H}{1} line profiles;
especially, reverberation mapping of these \ion{O}{1} and \ion{Ca}{2} lines would be a powerful tool to reveal the underlying physics.

\subsection{A Picture of the Dusty Emission Region}

The widely-accepted theory of the BELR describes its outer edge, where the \ion{O}{1} and \ion{Ca}{2} emission lines are likely to be formed,
set by the dust sublimation \citep[e.g.,][]{laor93, netzer93}.
If we accept this picture, the dust grains are possibly mixed in the line-emitting gas and suppress the \ion{Ca}{2} emission through the substantial
Ca depletion. 
Such a situation is in fact reported for the NELR by the absence or significant weakness of the observed [\ion{Ca}{2}] $\lambda$7291 line 
\citep{kingdon95, villar97}.
\citet{ferguson97} presented the LOC model calculations of the narrow emission lines and argued that Ca is depleted relative to the solar value
by factors of 3 -- 160.

It is hard to see an evidence of the dust presence in the emission region from our results, since Figure \ref{oioi} appears to show that our dust-free
models successfully reproduce the observations.
However, it is noteworthy that 
it is quite difficult to account for 
the observed data at \ion{O}{1} $n$($\lambda$ 11287)/$n$($\lambda$ 8446) $<$ 0.4 by the models with log $n_{\rm H}$ = 12.0.
The problem is that such small values of \ion{O}{1} $n$($\lambda$ 11287)/$n$($\lambda$ 8446) could only be reproduced with the higher densities than 
log $n_{\rm H}$ = 12.0 so that \ion{O}{1} $\lambda$8446 emission is exclusively enhanced by the collisional excitation, while such a dense gas produces 
intense \ion{Ca}{2} emission that is much stronger than observed.
One can clearly see this trend in Figure \ref{oioi} ({\it left}).
If we assumed significant Ca depletion in the line-emitting gas, these difficulties are naturally resolved since it significantly suppresses the 
otherwise intense \ion{Ca}{2} emissions.
For example, the data point representing (log $n_{\rm H}$, log $U$) = (13.0, $-$2.5) in Figure \ref{oioi} ({\it left}) could provide a plausible model for
the observations at \ion{O}{1} $n$($\lambda$ 11287)/$n$($\lambda$ 8446) $<$ 0.4 if the \ion{Ca}{2} emission is suppressed by a factor of a few times 10.\footnote{
The predicted EW for the model with (log $n_{\rm H}$, log $U$) = (13.0, $-$2.5) is EW (\ion{O}{1} $\lambda$8446) = 30 \AA, which is consistent with
the observations if rather large covering factors of 0.5 -- 1.0 and/or oxygen overabundance relative to the solar value are assumed.
}

The dust grains present in the line-emitting gas might also give the natural explanation to the observed lack of some UV emission lines relative to
their optical or near-IR counterparts.
\citet{ferland89} mentioned the possibility of the dust survival in the BELR gas in order to explain the extreme weakness of the observed \ion{Ca}{2}
$\lambda$3934 and $\lambda$3639 lines relative to the near-IR triplet.
At least a part of the long-standing \ion{Fe}{2} UV/opt problem, in which the observed ratios of \ion{Fe}{2} UV flux to the optical flux fall far below 
the photoionization model predictions \citep[see, e.g.,][]{baldwin04}, could also be explained.
However, it should be noted that the dust would affect the line formation processes in a very complicated manner, which should be precisely 
addressed when discussing the specific lines;
for the \ion{O}{1} and \ion{Ca}{2} lines, one of the most apparent effects as well as the Ca depletion would be the destruction 
of Ly$\beta$ photons which otherwise excite \ion{O}{1} atoms, thus the suppression of the \ion{O}{1} emissions.
The resultant line flux ratios could be much different from those derived from the simple speculations.


\section{Summary \label{sec:summary}}

We have compiled the emission-line fluxes of \ion{O}{1} $\lambda$8446, \ion{O}{1} $\lambda$11287, and the near-IR \ion{Ca}{2} triplet 
($\lambda$8579) observed in 11 quasars.
The quasars are distributed over wide ranges of redshift (0.06 $\le z \le$ 1.08) and of luminosity ($-29.8 \le M_B \le -22.1$), thus
providing a useful sample to track the line-emitting gas properties in various quasar environments.
The measured line strengths and velocities, as functions of the quasar properties, were analyzed with photoionization model calculations.
Our findings and conclusions are as follows:

1. There is no sign of a significant change in the flux ratios of the \ion{Ca}{2} triplet and \ion{O}{1} $\lambda$8446 over the
redshift and luminosity ranges studied here. 
It strongly indicates similar gas densities in the line-emission region from quasar to quasar.

2. The observed scatter of the \ion{O}{1} $\lambda$11287/$\lambda$8446 ratios appears to be related to the diversity of the
ionization parameter, while the ratio is not clearly dependent on the redshift or luminosity.
Combined with the similarity of the \ion{Ca}{2}/\ion{O}{1} $\lambda$8446 ratios,
it invokes the picture of the line-emitting gases in quasars that have similar densities and are located at regions exposed to various
ionizing radiation fluxes.

3. The \ion{O}{1} line widths are remarkably similar from quasar to quasar over more than 3 orders of magnitude in luminosity.
It might imply a kinematically determined location of the line emission region and is in clear contrast to the case of \ion{H}{1} lines,
whose emission region is considered to be radiation-selected.

4. If we accept that the \ion{O}{1} and \ion{Ca}{2} emission lines are formed at the outer edge of the BELR and that the outer edge is set by
the dust sublimation, the line-emitting gas is possibly mixed with the dust grains.
In fact such a situation may better reproduce the observations than the dust-free case through the significant Ca depletion.




\acknowledgments

We are grateful to the referee for giving useful and suggestive comments to improve this paper.
Y.M. acknowledges grants-in-aid from the Research Fellowships of the Japan Society for the Promotion of Science
(JSPS) for Young Scientists.
This work has been supported in part by Grants-in-Aid for Scientific Research
(15253002, 17104002) and the Japan-Australia Research Cooperative Program from JSPS.

\clearpage

\begin{table}
\begin{center}
\caption{Sample Characteristics\label{sample}}
\begin{tabular}{cccccccc}
\tableline\tableline
Quasar & Redshift & $M_B$\tablenotemark{a} & Reference\tablenotemark{b}\\
\tableline
QSO B0850$+$440    & 0.514 & $-$25.2 & 3\\
PG 1116$+$215      & 0.176 & $-$25.3 & 2\\
PG 1126$-$041      & 0.060 & $-$22.8 & 1\\
QSO J1139$-$1350   & 0.560 & $-$26.3 & 3\\
PG 1148$+$549      & 0.978 & $-$28.0 & 3\\
3C 273             & 0.158 & $-$26.9 & 3\\
PG 1415$+$451      & 0.114 & $-$22.1 & 1\\
PG 1448$+$273      & 0.065 & $-$23.0 & 1\\
PG 1519$+$226      & 0.137 & $-$22.9 & 1\\
PG 1612$+$261      & 0.131 & $-$22.6 & 1\\
PG 1718$+$481      & 1.084 & $-$29.8 & 3\\
\tableline
\end{tabular}
\tablenotetext{a}{The $B$-band absolute magnitude taken from \citet{veron06}.}
\tablenotetext{b}{References --- (1) \citet{riffel06}; (2) \citet{matsuoka05}; (3) Paper I}
\end{center}
\end{table}

\begin{table}
\begin{center}
\caption{Observed Photon Number Flux Ratios and EWs of the \ion{O}{1} and \ion{Ca}{2} Emission Lines \label{lineflux}}
\begin{tabular}{cccc}
\tableline\tableline
Quasar & $n$(\ion{Ca}{2})/$n$(\ion{O}{1} $\lambda$8446) 
& \ion{O}{1} $n$($\lambda$11287)/$n$($\lambda$8446) & EW (\ion{O}{1} $\lambda$8446) [\AA]\\
\tableline
QSO B0850$+$440    & 1.63 $\pm$ 0.53 & 0.62 $\pm$ 0.15 & 24.9 $\pm$ 5.9\\  
PG 1116$+$215      & 1.22 $\pm$ 0.43 & 0.67 $\pm$ 0.20 & 19.8 $\pm$ 5.2\\  
PG 1126$-$041      & 1.12 $\pm$ 0.06 & 0.52 $\pm$ 0.04 & 38.2 $\pm$ 1.4\\  
QSO J1139$-$1350   & 0.49 $\pm$ 0.16 & 0.24 $\pm$ 0.05 & 16.0 $\pm$ 2.9\\  
PG 1148$+$549      & 1.18 $\pm$ 0.05 & 0.33 $\pm$ 0.02 & 26.0 $\pm$ 0.9\\  
3C 273             & 1.46 $\pm$ 0.21 & 0.33 $\pm$ 0.04 & 17.2 $\pm$ 1.9\\  
PG 1415$+$451      & 0.87 $\pm$ 0.12 & 0.35 $\pm$ 0.05 & 23.3 $\pm$ 2.2\\  
PG 1448$+$273      & 0.34 $\pm$ 0.12 & 0.54 $\pm$ 0.08 & 22.1 $\pm$ 2.5\\  
PG 1519$+$226      & 0.97 $\pm$ 0.07 & 0.41 $\pm$ 0.03 & 31.9 $\pm$ 1.6\\  
PG 1612$+$261      & 0.47 $\pm$ 0.07 & 0.44 $\pm$ 0.05 & 35.5 $\pm$ 2.3\\  
PG 1718$+$481      & 1.11 $\pm$ 0.22 & 0.47 $\pm$ 0.12 &  7.6 $\pm$ 0.8\\  
\tableline
\end{tabular}
\tablenotetext{}{Note. --- The representative wavelength of 8579 \AA\ is used to convert
  the \ion{Ca}{2} triplet fluxes to the photon number fluxes.}
\end{center}
\end{table}

\clearpage



\end{document}